\documentclass{aa}
\input epsf
\begin{document}

   \thesaurus{08     
              (03.13.2;  
               08.02.3;  
               08.09.2 (Cyg X-1);  
               13.25.3)} 
   \title{Frequency resolved spectroscopy of Cyg X-1: fast variability
   of the Fe $K_{\alpha}$ line.}

   \author{M. Revnivtsev\inst{1,2}, M. Gilfanov\inst{2,1},
   E. Churazov\inst{2,1}
	}

   \offprints{revnivtsev@hea.iki.rssi.ru}

   \institute{Space Research Institute, Russian Academy of Sciences,
              Profsoyuznaya 84/32, 117810 Moscow, Russia,
        \and
                Max-Planck-Institute f\"ur Astrophysik,
              Karl-Schwarzschild-Str. 1, 85740 Garching bei M\"unchen,
              Germany
             }
  \date{}

	\authorrunning{Revnivtsev M. et al.}
	\titlerunning{Frequency resolved spectroscopy: Cyg X-1}
	
   \maketitle

\sloppypar

   \begin{abstract}

We studied the frequency resolved energy spectra of Cyg X--1 during the
standard low (hard) spectral state using the data of the Rossi X-Ray
Timing Explorer. We found that the relative amplitude of the
reflection features -- the iron fluorescent line at $\sim 6.5$ keV and
the smeared edge above $\sim 7$ keV -- decreases with the increasing
frequency. In particular we found that the equivalent width of the
iron line decreases above $\sim 1$ Hz and drops twice at frequency of $\sim
10 $ Hz.

An assumption that such behavior is solely due to a finite light crossing
time of the reflecting  media, would imply the characteristic size of
the reflector $\sim 5\times10^8$ cm,    corresponding to $\sim 150
R_{\rm g}$ for a $10M_{\sun}$ black hole.  Alternatively lack of high
frequency oscillations of the reflected component may indicate  that
the short time scale, $\sim$ 50--100 msec,  variations of the primary
continuum  appear  in geometrically different, likely inner, part of
the accretion flow and give a rise to a significantly weaker, if any,
reflected emission than the longer time scale events.

   \end{abstract}

%

\section{Introduction}

Over the last several years it has become widely accepted that the
Galactic X-ray binaries exhibit Compton reflection features in
their spectra. 

Qualitatively, the reflection component in the X-ray
spectra of these systems can be described as a broad ``hump'' at  
energies 20--30 keV, fluorescent Fe line at energies
6.4--6.7 keV (depending on the ionization  state of  the reflecting
medium) and an absorption edge at the energy $\approx$7.1 keV
(\cite{shef}, \cite{fab}). For all geometries which do not 
obscure the direct primary continuum from the observer, the detected
energy spectra consists of this direct component (approximately a power
law in the 3--13 keV energy band) and the reflection features.   
For standard  cosmic abundances the    expected equivalent width of the
fluorescent Fe line is $\sim$100--200 eV  (for a source above semi--infinite slab of
neutral matter). 

 The geometry and mutual location of the source of primary continuum and
reflecting medium should affect both the equivalent width of the
fluorescent line and the character of it's variability. In particular the
finite size of the reflector implies that the time variations of the
reflected radiation should be smeared out on the time scales corresponding
to the light crossing time of the reflector. In addition, a time
lags between different emission components might apprear.
  Alternatively the geometrically
different region of the main energy release zone (where the primary
continuum is produced) may have different efficiency for production of the
reflection component. As a result timing properties of the reprocessed 
component may be linked to the properties of the selected region of this zone.

\section{Observations and data analysis}

For our analysis we chose publicly available data of the Rossi
X-ray Timing Explorer observations P10238 performed between Mar. 26, 1996 and
Mar. 31, 1996 with a total exposure time $\sim$70 ksec (we used  only 
observations were all 5 PCU were turned on).
For the frequency resolved spectral analysis we used PCA data in
the ``Generic Binned'' mode, having $\frac{1}{64}$sec$\sim$16 msec time 
resolution in 64 energy channels covering the whole PCA energy band
(B\_16ms\_64M\_0\_249). For analysis of the averaged spectrum we used
the ``Standard Mode 2'' data, having twice as many energy channels in the
energy band of our interest (3--13 keV) than the ``Generic Binned'' mode. 

For the obtained spectra we constructed the response matrix using the
standard tasks of FTOOLS 4.2 package with PCA RMF v3.5 (Jahoda 1999). The
comparison of the timing mode spectra with that obtained from ``Standard
Mode 2'' showed, that they do not differ more than by 1--1.5\%. The
background was calculated using ``Q6'' model ($VLE$-based model,
preferable for bright sources does not work for these 
observations). However, the PCA background is negligible for the time
averaged spectra of Cyg X-1 in the energy band 3--13 keV. The PCA 
background contribution to the frequency resolved spectra at the
frequencies 0.03--30 Hz is even less important.

\section{Frequency resolved spectral analysis}

We calculated the Power Density Spectra (PDSs) in each energy channel for
every 128 sec time segments using the standard Fast Fourier Transform
(FFT) procedure and adopting the normalization of Miyamoto et al. (1991): 

$$ P_j=2|a_j|^2/N_{\gamma}R$$
$$a_j=\sum_{k=1}^{2^m}{x_k e^{i\omega_jt_k}}$$

\noindent
where $t_k$ is the time label of the $k$th bin, $x_k$ is the number
of counts in this bin, $N_{\gamma}$ is the total number of photons and
$R$ is the average count rate in the segment 
($R=N_{\gamma}/T$, $T$ -- total duration of the light curve
segment). This PDS normalization ((rms/mean)$^2$/Hz) allows one to obtain
squared fractional $rms$ by integrating the power over the frequencies
of interest. Then the obtained power spectra were averaged over
the light curve segments and the logarithmic frequency bins.
For every obtained frequency bin the frequency dependent spectra were
constructed according to the formula:

$$ S(E_i,f_j)=R_i\sqrt{P_i(f_j)\Delta f_j}=\sqrt{{{2|a_{ij}|^2}\over{T}}\Delta f_j} $$

\noindent
here $S(E_i,f_j)$ is the count rate of the spectrum on the frequency
$f_j$ in the energy channel $E_i$. Since the light curves of Cyg X-1
in the different energy bands almost perfectly coherent and have 
practically the same form of the PDS (see e.g. Nowak et al. 1999), the
spectra obtained using the procedure above can be used to 
describe the real energy spectrum of the source X-ray flux variations
on the different time scales. The above expession also can be easily
adapted to describe the energy spectrum of different PDS
spectrum components.

\begin{figure}
\epsfxsize 8 cm
\epsffile[50 180 560 710]{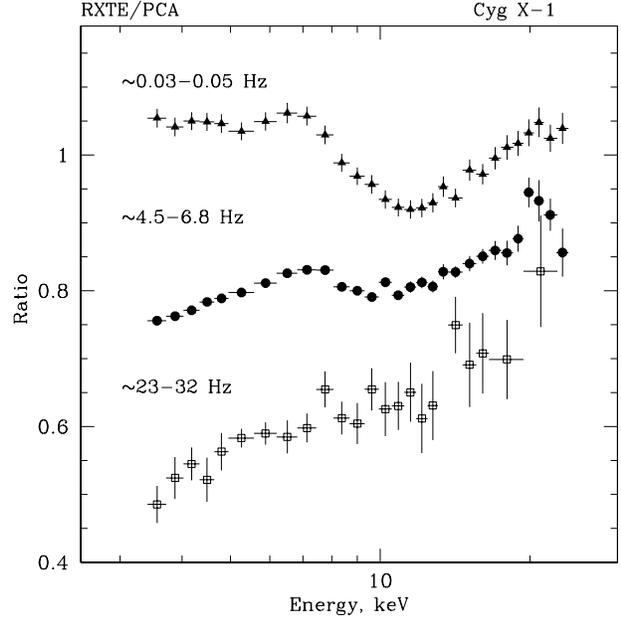}
\caption{The ratio of the energy spectra of Cyg X-1 in different frequency
bands to a power law model with photon index $\alpha=1.8$. Spectra
corresponding to 0.03-0.05 Hz and 23--32 Hz were rescaled for clarity.
\label{spectra}
}
\end{figure}

\begin{figure}
\epsfxsize 8 cm
\epsffile[50 190 560 710]{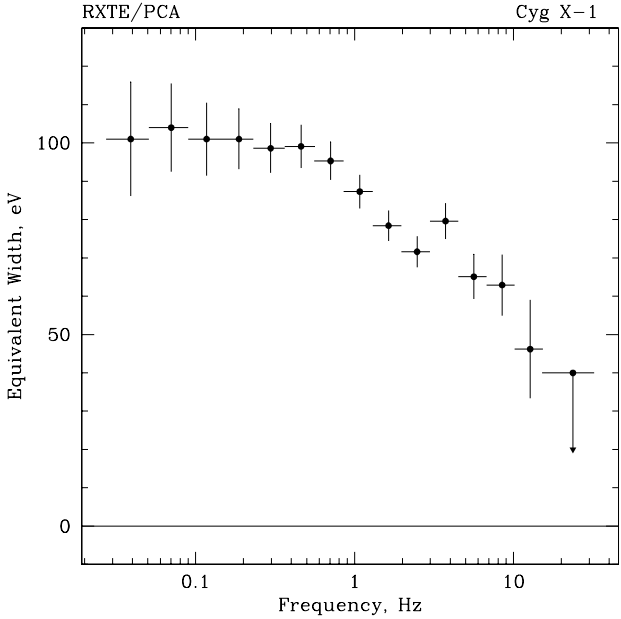}
\caption{Dependence of the equivalent width of the fluorescent Fe
line on the frequency. For the spectral approximation the
powerlaw+gaussian line model was used (3--13 keV energy band, the
centroid energy and the width of the line were frozen at the values 6.4
keV and 0.1 keV respectively).
\label{eqw}
}
\end{figure}

\section{Results}

The energy spectra of Cyg X--1 in several narrow frequency ranges
between 0.03 and 30 Hz are shown in Fig. \ref{spectra}. 
The change of the spectral shape with the decrease of the characteristic
time scale  can be clearly seen. 
The relative amplitude of the reflection features, in particular that
of the broad line at $\sim$ 6--7 keV and the ``smeared edge'' above
$\sim 7$ keV, is apparently decreasing with the increase of frequency and
vanishes above $\sim 20$ Hz. Along with becoming less
``wiggled'' the spectra become harder as frequency increases.

In order to quantify these effects we fit frequency resolved spectra in
16 frequency bins covering 0.03--32 Hz range
in the 3--13 keV energy band with a simple two component model
consisting of a power law and a Gaussian line with the fixed energy
$E_{\rm line}=6.4$ keV and the width FWHM=0.1 keV (that is comparable
with the PCA energy resolution at this energy). 
The best fit parameters are given in Table \ref{params}.
The dependence of the
line equivalent width upon frequency is shown in Fig. \ref{eqw}.
As is seen from the Fig. \ref{eqw}, the equivalent width of the line 
starts decreasing above $\sim 1 $ Hz and falls by a factor of 2 at
$\sim 10$ Hz. No significant flux in the line was detected above 15 Hz
with the 2$\sigma$ upper limit on the equivalent width in the
15--32 Hz frequency range of $\approx$40 eV.

Variations of the parameters of the
spectral model, in particular the change of the line centroid to 6.0
and 6.7 keV  and increase of the line width to 1.0 keV, do not
change the general trend. These variations, however, affect the
particular values of the equivalent width and, to some extent, the
shape of the curve in Fig. \ref{eqw}. The change of the line
centroid energy to the 5 keV or 9 keV, on the contrary, results in
disappearance of the effect.

Finally we should mention that similar behaviour of the frequency
resolved spectra was detected in Cyg X-1 observations 20175-01-01-00,
20175-01-02-00 and in all available observations of the set P30157
(01--10), performed in 1997 but with lower significance.

\begin{table}
\small
\caption{Best fit parameters of the approximation of frequency dependent
spectra of Cyg X-1 with the power law + gaussian line (line energy and 
width were frozen at the values 6.4 keV and 0.1 keV respectively) in
the energy band 3--13 keV. 
\label{params}}  

\begin{tabular}{lcc}
\hline
Freq., Hz&PL slope,$ \alpha$&EW, eV\\
\hline
$(3.9\pm1.2)\times10^{-2}$&$1.93\pm0.01$&$101\pm15$\\
$(7.0\pm1.9)\times10^{-2}$&$1.91\pm0.01$&$104\pm 12$\\
$0.12\pm0.03$&$1.91\pm0.01$&$101\pm 9$\\
$0.19\pm0.04$&$1.89\pm0.01$&$101\pm 8$\\
$0.30\pm0.07$&$1.89\pm0.01$&$98.6\pm 6.4$\\
$0.46\pm0.09$&$1.89\pm0.01$&$99.1\pm 5.6$\\
$0.7\pm0.14$&$1.89\pm0.01$&$95.3\pm 4.9$\\
$1.07\pm0.22$&$1.86\pm0.01$&$87.3\pm 4.4$\\
$1.63\pm0.33$&$1.83\pm0.01$&$78.4\pm 3.9$\\
$2.48\pm0.50$&$1.81\pm0.01$&$71.6\pm 4.0$\\
$3.74\pm0.76$&$1.80\pm0.01$&$79.6\pm 4.7$\\
$5.63\pm1.14$&$1.75\pm0.01$&$65.1\pm 5.8$\\
$8.48\pm1.71$&$1.70\pm0.01$&$62.9\pm 7.9$\\
$12.7\pm2.6$&$1.63\pm0.01$&$46.2\pm 12.8$\\
$19.2\pm3.8$&$1.62\pm0.01$&$<48^a$\\
$27.5\pm4.5$&$1.63\pm0.01$&$<57^a$\\
\hline
\end{tabular}

$^a$ -- $2\sigma$ upper limit
\end{table}

\section{Discussion}

A time averaged energy spectrum of Cyg X-1 (e.g. \cite{ebisawa}, \cite{zdz}) as well as
frequency resolved spectra at low frequency 
(Fig.\ref{spectra}) show clear deviation from a single slope power law
spectrum. This deviation is commonly ascribed to the reflection from an
optically thick media located in the vicinity of the
production site of primary hard X--ray radiation. In a simple case of
reflection from an optically thick  cold neutral medium the main
reflection features   are well known
(\cite{shef}, \cite{fab}) -- a narrow unshifted iron 
$K_{\alpha}$ line at 6.4 keV with an equivalent width of $\sim 100$ eV,
an absorption edge at $\approx 7.1$ keV 
and a reflected continuum peaked at $\sim 20-30$ keV. However, the
real spectra of compact X--ray binaries show considerably more complicated
behavior. In particular the centroid energy of the line is often
different from 6.4 keV, the line width in many cases is as large as
$\sim 1$ keV and a broad ``smeared edge'' above $\sim$ 7--8 keV is
observed instead of a sharp absorption edge at $\approx 7.1$ keV
(\cite{ebisawa}).  
This departure from a simple reflection model hints at a
complicated ionization state and/or motion (e.g. Keplerian motion in
the disk) of the reflecting media.

It is obvious, that the simple spectral model used for fitting the
data in the previous section is
neither adequate nor completely  justified  from the physical point
of view. The apparent line centroid energy and width vary
with frequency. In particular the best fit centroid energy shifts
below 6.0 keV and the  width of the line exceeds 1 keV  as
frequency increases. 
However, due to the lack of a satisfactory realistic model of reflection
in an X--ray binary, especially applicable to the frequency
resolved spectral data,  we restricted ourselves to the simple
two component model described in the previous section. The main
purpose of using this spectral model was to quantify the major effect 
which is clearly seen in Fig.\ref{spectra}, namely the suppression of the
reflection features in the energy spectrum with the increase of
frequency. 

  The results of the analysis presented in this paper show that 
the relative amplitude of the reflected component variations is lower on the 
short time scales ($\la$ 50--100 msec), than that of the primary X-ray
radiation. Similar effect was found Oosterbroek et al. (1996) for GS
2023+338 but the characteristic time scale in this case was much
longer, $\sim$200 sec. GS 2023+338 is peculiar in many ways.
In particular the source exhibited strong ($\ga 10^{23}$ cm$^{-2}$)
and highly variable on the time scales of $\sim 10^3$ sec low energy
absorption (e.g. \cite{done}) and, most importantly, extremely
strong Fe line with equivalent width up to 1.4  keV (\cite{ooster}),
that could mean that  geometry of 
the reprocessing media in GS2023+338 is different from that in Cyg X--1.
Recently we found suppression of the reflected component variations in
the low spectral state of another black hole candidate GX339--4
(Revnivtsev, Gilfanov \& Churazov, 1999). In this case the
characteristic time scale ($\sim$50--100 msec) was similar to that in Cyg X--1.
We therefore can tentatively conclude that such a behavior might be a
common feature of the black hole binaries in the low spectral state.

A most straightforward explanation of this effect  would be in
terms of a finite light crossing time of the reflector $\tau_{\rm refl}\sim
l_{\rm refl}/c$, caused by a finite spatial  extend $l_{\rm refl}$  of the
reflecting media. From Fig.\ref{eqw} one can see that the equivalent width
of the iron line drops by a factor of two at the frequency $f_{1/2}\sim 10$
Hz. This value gives us a rough  estimate of the characteristic response
time of the reflector $\tau_{\rm refl}\sim 1/2\pi f_{1/2}\sim 15$ msec and the
characteristic  size of the reflecting media $l_{\rm refl}\sim 5\times
10^8$ cm which would correspond to $\sim 150 R_{\rm g}$ for a $10M_{\sun}$
black hole. We note however that if the primary continuum originates within
the $\sim 10 R_{\rm g}$ sphere in the inner zone of the accretion disk,
then only  a small fraction of the emitted hard radiation has the
possibility to be reflected from the accretion disk regions with $R\ga
150R_{\rm g}$ (in the case of a flat disk) and it can hardly provide observed
$\sim 100$ eV equivalent width of the iron line. Therefore the
assumption that high frequency variations of the reflected component
are caused by the finite light crossing time of the reflector implies
interesting constraints on the geometry of the source of primary
continuum and reflector.

An alternative explanation might be that the short time scale, $\la$
50--100 msec, variations appear in geometrically different, likely
inner, part of the accretion flow and give a rise to significantly
weaker, if any, reflected emission than the longer time scale events,
presumably originating in the outer regions. This might be caused, for
instance,  by a smaller solid angle of the reflector as seen by the short
time scale events and/or due to screening of the reflector from  the short
time scale events by the outer parts of the accretion flow.

\begin{acknowledgements}
This research has made use of data obtained through the High Energy
Astrophysics Science Archive Research Center Online Service, provided 
by the NASA/Goddard Space Flight Center.
\end{acknowledgements}

\end{document}